\documentclass[english,pra,notitlepage,preprintnumbers]{revtex4-1}
\usepackage[latin9]{inputenc}
\usepackage{geometry}
\usepackage{amsmath}
\usepackage{amssymb}
\usepackage{graphicx}
\usepackage{babel}
\begin{document}
\global\long\def\calM{\mathcal{M}}%
\global\long\def\me{m}%
\global\long\def\MP{M}%

\preprint{Alberta Thy 6-22}
\title{Positronium hydride decay into proton, electron, and one or zero photons}

\author{M. Jamil Aslam}
\affiliation{Department of Physics, University of Alberta, Edmonton,
  Alberta, Canada T6G 2E1}
\affiliation{Department of Physics, Quaid-i-Azam University, Islamabad, Pakistan}
\author{Wen Chen}
\altaffiliation[Current address:]{School of Physics,  Zhejiang 
  University,  Hangzhou,  Zhejiang 310027,  China}
\affiliation{Department of Physics, University of Alberta, Edmonton,
  Alberta, Canada T6G 2E1}

\author{Andrzej Czarnecki}
\author{Muhammad Mubasher}
\author{Connor Stephens}
\affiliation{Department of Physics, University of Alberta, Edmonton,
  Alberta, Canada T6G 2E1}
\maketitle
Decay rates of the positronium hydride PsH, a bound state of a proton,
a positron, and two electrons, are determined for two rare channels,
$\text{PsH}\to p^{+}e^{-}\gamma$ and $\text{PsH}\to p^{+}e^{-}$.
Previous studies overestimated these rates by factors of about 2 and
700, respectively. We explain the physics underlying these wrong predictions.
We confirm a range of static PsH properties, including the non-relativistic
ground state energy, expectation values of inter-particle distances
and their powers, and the three and four particle coalescence probabilities,
using a variational method in the Gaussian basis. 

\section{Introduction}

Positronium hydride (PsH) consists of a proton $p^{+}$, positron
$e^{+}$, and two electrons $e^{-}$. Stable with respect to autoionization,
it decays due to electron-positron annihilation. Similarly to the
case of the positronium-ion ($\text{Ps}$$^{-}$), its two electrons
form a spin singlet. When the positron and one of the electrons meet,
they can form a spin singlet or a triplet. Their annihilation can
lead to final states with any number of photons, even or odd. Here
we calculate the rate of decays that result in one or no photons,
as well as unbound electron and proton. This is the first calculation
of these decay rates. Previously they were only estimated and we explain
why those estimates were incorrect. The key issue is the role of the
proton in influencing the $e^{+}e^{-}$ annihilation.

In addition, we re-evaluate the wave function of PsH using the variational
method with a Gaussian basis. In order to test it, we calculate the
non-relativistic ground state energy, mean inter-particle distances,
and, most importantly for our purposes, probabilities of coalescence
of $e^{+}e^{-}e^{-}$ and of $p^{+}e^{+}e^{-}e^{-}$. We confirm the
values of these quantities found in Ref.~\citep{Bubin11}. 

The paper is organized as follows: In Section \ref{sec:Brief-history}
we put our study in the context of previous work on PsH. In Section
\ref{sec:PsH-WF}, we discuss its Hamiltonian and wave function. Section
\ref{sec:One-photon-decay} focuses on the decay $\text{PsH}\to p^{+}e^{-}\gamma$
and Section \ref{sec:Zero-photon-decay} on $\text{PsH}\to p^{+}e^{-}$.
We conclude by comparing our results with previous literature in Section
\ref{sec:Conclusions}. 

We use such units that $\hbar=c=\epsilon_{0}=1$, except for the expectation
values of operators computed with the variational wave function of
PsH, given in atomic units, as explained in Section \ref{sec:PsH-WF}.
We denote the electron mass by $\me$ and the proton mass by $M$.
Unless indicated otherwise, we neglect the binding energy of PsH in
comparison with $\me$ and treat its constituents as stationary particles,
neglecting their relative motion. Corrections to this approximation
are suppressed by the fine structure constant $\alpha\simeq1/137$. 

\section{Brief history of $\text{PsH}$\label{sec:Brief-history}}

In their pioneering works, Wheeler \citep{Wheeler:1946xth} and Hylleraas
and Ore \citep{Hylleraas:1947zza} studied small exotic molecules
where one or more nuclei are replaced by positrons. Ore \citep{Ore:1951zz}
established the stability of the PsH ground state. Since then, much
theoretical work has been done on the energy of ground, metastable,
and resonant states and on other properties of this system (see for
example \citep{Neamtan1962,goldanskii:1967aa,Lebeda:1969,Navin1974,Ho:1978,Yoshida:1996,Frolov:1997aa,Nagashima:1998}
where further references can be found).

Experimental efforts to produce and detect this system have also been
made. Pareja et al.~\citep{Pareja90} first reported the existence
of such a bound state in a condensed phase. Further evidence was provided
by Schrader et al.~\citep{Schrader92} in positron-methane collisions,
\begin{equation}
e^{+}+\text{CH}_{4}\to\text{CH}_{3}^{+}+\text{PsH},
\end{equation}
with an estimated binding energy, $E_{b}=-1.1\pm0.2$ eV, in line
with most theoretical predictions.  

PsH is a special case of a Coulombic system, positioned between the
hydrogen molecule H$_{2}$ and dipositronium Ps$_{2}$,
in which both nuclei are replaced with positrons. Since positron's
motion cannot be considered as slow,
$\text{PsH}$ is an essentially four body system.

On the theoretical side, exotic systems containing antimatter
serve to test various quantum mechanical methods. Over the years,
the accuracy of theoretical calculations in $\text{PsH}$ has improved
thanks to advances in computational techniques and increased hardware
power. Using variational methods to obtain accurate wave functions,
most of the studies performed for the ground state energy of such
system are non-relativistic; relativistic effects have been calculated
by Yan and Ho \citep{Yan99} and by Bubin and Varga \citep{Bubin11}. 

An interesting problem is the study of electron-positron annihilation
in PsH producing zero, one, two, and in general $n$ photons. What
makes it more interesting is that the electron and proton can either be
free or form a bound hydrogen state. Refs.~\citep{Houston73,Page1974,NavinPRA74}
considered both bound and unbound final states. In case of unbound
electron and $p^{+}$ final states, estimates were given for the two
photon annihilation rate $\Gamma_{2\gamma}=\Gamma\left(\text{PsH}\to p^{+}e^{-}\gamma\gamma\right)$,
the dominant process. The rate of annihilation into three or more photons
$\left(\text{\ensuremath{\Gamma_{n\gamma}}, \ensuremath{n\ge3}}\right)$
can be found using $\Gamma_{2\gamma}$ and the rate of the $n\gamma$
decay in a positronium atom. This is the subject of Ferrante relations
\citep{Ferrante:1968zz}, justified in \citep{Frolov:1997hy}. 

Decays with one or no photons have been estimated using analogous 
$\text{Ps}^{-}$  and $\text{Ps}_{2}$
results \citep{Frolov:1997hy,FROLOV2005430,Frolov:2009qi} in the
absence of 
a dedicated QED calculation for PsH. Filling
this gap is the main motivation of this paper. 

\section{$\text{PsH}$ wave function and Hamiltonian\label{sec:PsH-WF}}

We label coordinates of the proton with 1, positron with 2, and 
electrons with 3 and 4.
The $\text{PsH}$ wave function is a product of spatial and spin parts,
antisymmetrized with respect to permuting the electrons,
\begin{equation}
\psi=\chi_{\uparrow}^{2}\left(\chi_{\downarrow}^{3}\chi_{\uparrow}^{4}-\chi_{\uparrow}^{3}\chi_{\downarrow}^{4}\right)\left(1+P_{34}\right)\phi_{S}\label{eq:wf}
\end{equation}
where $\chi$s denote spin states; $P_{34}$ is the permutation
operator of the electrons; and $\phi_{S}$ is the S-wave
spatial wave function. In the Gaussian basis \citep{Puchalski:2008jj},
that spatial part is written as
\begin{equation}
\phi_{S}=\sum_{i=1}^{N}c_{i}^{S}\exp\left[-\sum_{a<b}w_{ab}^{iS}r_{ab}^{2}\right]\label{eq:swf}
\end{equation}
where $w_{ab}$ are real coefficients and $N$ is the number of trial
functions (basis size). Factors of $1/\sqrt{2}$ from the
permutation operator and $1/\sqrt{4\pi}$ from the S-state
wave function are absorbed in the normalization of linear coefficients
$c_{i}^{S}.$

The proton, much heavier than the remaining constituents, is
sometimes treated as a static source of the electric field \citep{Ho1986}.
In our approach, we follow the analogy with dipositronium
\cite{Puchalski:2008jj} and include the motion of all four bodies.
However, we neglect the magnetic moment of the proton throughout this
paper so that the spin of the positron is the total angular momentum
of PsH, a constant. The Coulomb Hamiltonian is
\begin{align}
\hat{H} & =\sum_{i=1}^{4}\frac{\hat{p}_{i}^{2}}{2m_{i}}+\sum_{i<j}V\left(r_{ij}\right)\nonumber \\
 & =\frac{\hat{p}_{1}^{2}}{2m_{1}}+\frac{\hat{p}_{2}^{2}}{2m_{2}}+\frac{\hat{p}_{3}^{2}}{2m_{3}}+\frac{\hat{p}_{4}^{2}}{2m_{4}}+\alpha\sum_{i<j}\frac{z_{i}z_{j}}{r_{ij}},\label{eq:1}
\end{align}
where $z_{i}$ equals $-1$ for $e^{-}$ and $+1$ for $e^{+}$ and
$p^{+}$. Electron and positron masses are denoted by $m_{2}=m_{3}=m_{4}\equiv\me$.
In atomic units (a.u.) we take $\me=1$ and $m_{1}=\MP\simeq1836$. 

Let $\vec{A}_{i}$ denote the absolute coordinates and $\vec{r}_{ij}$
the relative coordinates. The inter-particle distances are $r_{ij}=\sqrt{\left(\vec{A}_{i}-\vec{A}_{j}\right)^{2}}$. In terms
of these coordinates, the Hamiltonian (\ref{eq:1}) becomes
\begin{align}
\hat{H} & =-\frac{1}{2\mu_{12}}\left[\vec{\nabla}_{\vec{r}_{12}}^{2}+\vec{\nabla}_{\vec{r}_{13}}^{2}+\vec{\nabla}_{\vec{r}_{14}}^{2}\right]-\frac{1}{m_{1}}\left[\vec{\nabla}_{\vec{r}_{12}}\cdot\vec{\nabla}_{\vec{r}_{13}}+\vec{\nabla}_{\vec{r}_{12}}\cdot\vec{\nabla}_{\vec{r}_{14}}+\vec{\nabla}_{\vec{r}_{13}}\cdot\vec{\nabla}_{\vec{r}_{14}}\right]\nonumber \\
 & +\alpha\left[\frac{z_{1}z_{2}}{r_{12}}+\frac{z_{3}z_{4}}{r_{34}}+\frac{z_{1}z_{3}}{r_{13}}+\frac{z_{1}z_{4}}{r_{14}}+\frac{z_{2}z_{3}}{r_{23}}+\frac{z_{2}z_{4}}{r_{24}}\right].\label{eq:Hamiltonian}
\end{align}
where $\mu_{ij}=\frac{m_{i}m_{j}}{m_{i}+m_{j}}$ is the reduced mass
and in our case $\mu_{12}=\mu_{13}=\mu_{14}$. Translating from  absolute
to relative coordinates, we have ignored the kinetic energy of the
centre-of-mass motion of the $\text{PsH}$ system. 

\def\refBubin{\text{Ref.~\citep{Bubin11}}}  
\begin{table}[h]
\[
\begin{array}{ccccccc}
 &  & \left\langle r_{p^{+}e^{+}}\right\rangle  & \left\langle r_{e^{+}e^{-}}\right\rangle  & \left\langle r_{p^{+}e^{-}}\right\rangle  & \left\langle r_{e^{-}e^{-}}\right\rangle  & \left\langle r_{p^{+}e^{+}}^{2}\right\rangle \\
 & \text{Bubin } & 3.663\;50 & 3.481\;18 & 2.313\;16 & 3.577\;0 & 16.272\\
 & \text{Ours} & 3.663\;47 & 3.481\;16 & 2.313\;15 & 3.577\;0 & 16.272\\
\\
 &  & \left\langle r_{e^{+}e^{-}}^{2}\right\rangle  & \left\langle r_{p^{+}e^{-}}^{2}\right\rangle  & \left\langle r_{e^{-}e^{-}}^{2}\right\rangle  & \left\langle 1/r_{p^{+}e^{+}}^{2}\right\rangle  & \left\langle 1/r_{e^{+}e^{-}}^{2}\right\rangle \\
 & \refBubin & 15.593\;54 & 7.824\;79 & 15.895\;94 & 0.172\;0 & 0.349\\
 & \text{Ours} & 15.593\;22 & 7.824\;54 & 15.895\;43 & 0.172\;0 & 0.349\\
\\
 &  & \left\langle 1/r_{p^{+}e^{-}}^{2}\right\rangle  & \left\langle 1/r_{e^{-}e^{-}}^{2}\right\rangle  & \left\langle 1/r_{p^{+}e^{+}}\right\rangle  & \left\langle 1/r_{e^{+}e^{-}}\right\rangle  & \left\langle 1/r_{p^{+}e^{-}}\right\rangle \\
 & \refBubin & 1.205\;65 & 0.213\;65 & 0.347\;30 & 0.418\;43 & 0.729\\
 & \text{Ours} & 1.205\;62 & 0.213\;65 & 0.347\;30 & 0.418\;43 & 0.729\\
\\
 &  & \left\langle 1/r_{e^{-}e^{-}}\right\rangle  & \left\langle
                                                    T\right\rangle  &
                                                                      \left\langle V\right\rangle  & \left\langle \hat H\right\rangle  & \left\{ \begin{array}{c}
\left\langle \delta_{e^{+}e_{3}^{-}}\delta_{e^{+}e_{4}^{-}}\right\rangle \\
\equiv\left\langle \delta_{e^{+}e_{3}^{-}}\delta_{e_{3}^{-}e_{4}^{-}}\right\rangle 
\end{array}\right.\\
 & \refBubin & 0.370\;33 & -- & -- & -0.788\;87 & 3.7147\times10^{-4}\\
 & \text{Ours} & 0.370\;33 & 0.788\;87 & -1.577\;74 & -0.788\;87 & 3.7364\times10^{-4}\\
\\
 &  & \left\{ \begin{array}{c}
\left\langle \delta_{p^{+}e^{+}}\delta_{p^{+}e^{-}}\right\rangle \\
\equiv\left\langle \delta_{p^{+}e^{+}}\delta_{e^{+}e^{-}}\right\rangle 
\end{array}\right. & \left\langle \delta_{p^{+}e^{+}}\delta_{e_{3}^{-}e_{4}^{-}}\right\rangle  & \left\langle \delta_{p^{+}e_{3}^{-}}\delta_{e^{+}e_{4}^{-}}\right\rangle  & \left\langle \delta_{p^{+}e_{3}^{-}}\delta_{p^{+}e_{4}^{-}}\right\rangle  & \left\langle \delta_{p^{+}e^{+}}\delta_{p^{+}e_{3}^{-}}\delta_{e^{+}e_{4}^{-}}\right\rangle \\
 & \refBubin & 8.6\times10^{-4} & 3.16\times10^{-5} & 6.32\times10^{-3} & 7.5334\times10^{-3} & 1.9038\times10^{-4}\\
 & \text{Ours} & 8.8\times10^{-4} & 3.12\times10^{-5} & 6.09\times10^{-3} & 7.3087\times10^{-3} & 1.8018\times10^{-4}\\
\\
\end{array}
\]

\caption{Values of physical parameters for the $\text{PsH}$ calculated using
Gaussian wave functions, compared with results of Ref.~\citep{Bubin11}.
All values are given in atomic units where the unit of length is the
Bohr radius $\hbar/\left(\alpha mc\right)$. Basis size is always
1000.\label{tab:Values-of-physical}}
\end{table}
 
The expectation value of  the Hamiltonian
with the wave function in Eq.~(\ref{eq:swf})
approximates 
the ground state energy
 in terms of six exponents
$w_{ab}^{iS}$. These six parameters are determined, for each of the
$N$ elements of the basis, following the optimization method described
in \citep{Puchalski:2008jj}. The results for a range of parameters
of the $\text{PsH}$ system are given in Table \ref{tab:Values-of-physical}
along with the corresponding values calculated in Ref.~\citep{Bubin11}.
We find good agreement, especially for the non-relativistic ground
state energy $\left\langle \hat H\right\rangle$. The binding energy (dissociation energy) is
(in atomic units, taking $\alpha^{2}mc^{2}$ as the unit energy)
\begin{align}
E_{b} & =-\left\langle \hat H\right\rangle +E^{\text{H}}+E^{\text{Ps}}\nonumber \\
 & =-\left\langle \hat H\right\rangle -\frac{3}{4}\text{a.u.},\label{eq:binding}
\end{align}
where $\left\langle \hat H\right\rangle$ is given in Table \ref{tab:Values-of-physical} and
the ground state energies of hydrogen and positronium are $-\frac{1}{2}\text{a.u.}$
and $-\frac{1}{4}\text{a.u.}$, respectively. The results we will
use in Sections \ref{sec:One-photon-decay} and \ref{sec:Zero-photon-decay}
are 
\begin{align}
\left\langle \delta_{e^{+}e_{3}^{-}}\delta_{e_{3}^{-}e_{4}^{-}}\right\rangle  & =3.73\left(2\right)\cdot10^{-4},\label{eq:coal3}\\
\left\langle \delta_{p^{+}e^{+}}\delta_{p^{+}e_{3}^{-}}\delta_{e^{+}e_{4}^{-}}\right\rangle  & =1.85\left(1\right)\cdot10^{-4}.\label{eq:coal4}
\end{align}
The central values are arithmetic means of the results in Ref.~\citep{Bubin11}
and ours. Their differences are used as error estimates. 

\section{One-photon decay $\text{PsH}\to p^{+}e^{-}\gamma$\label{sec:One-photon-decay}}

\begin{figure}[h]
\centering\includegraphics[scale=0.4]{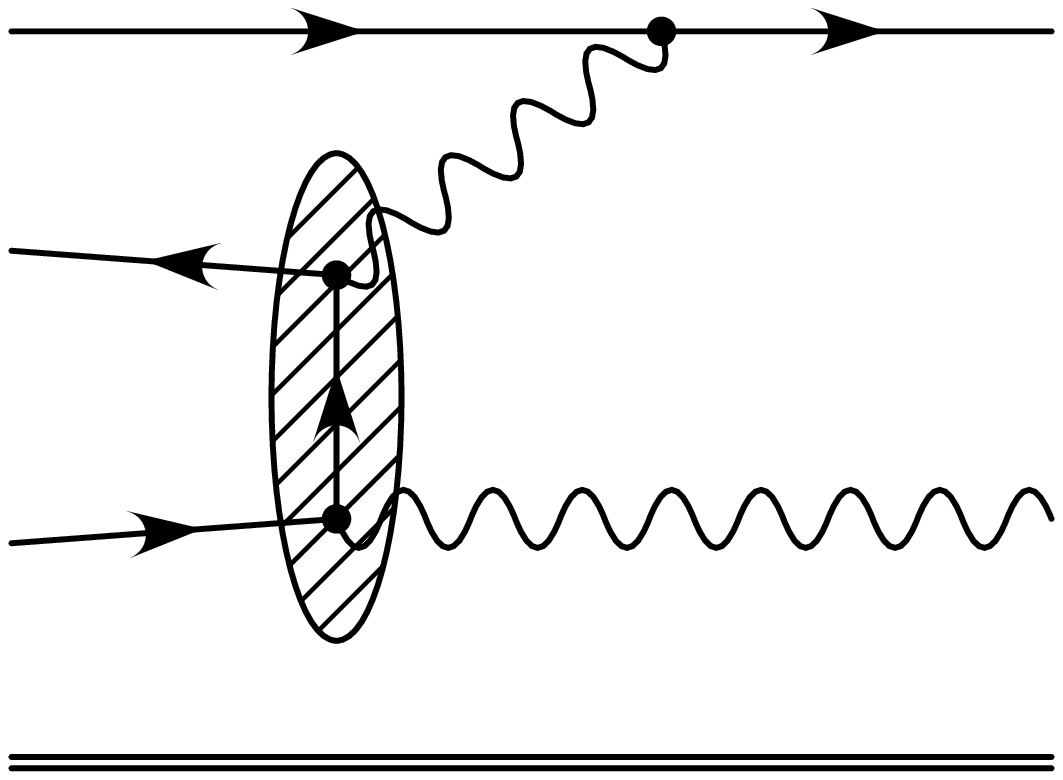}\hspace*{10mm}\includegraphics[scale=0.4]{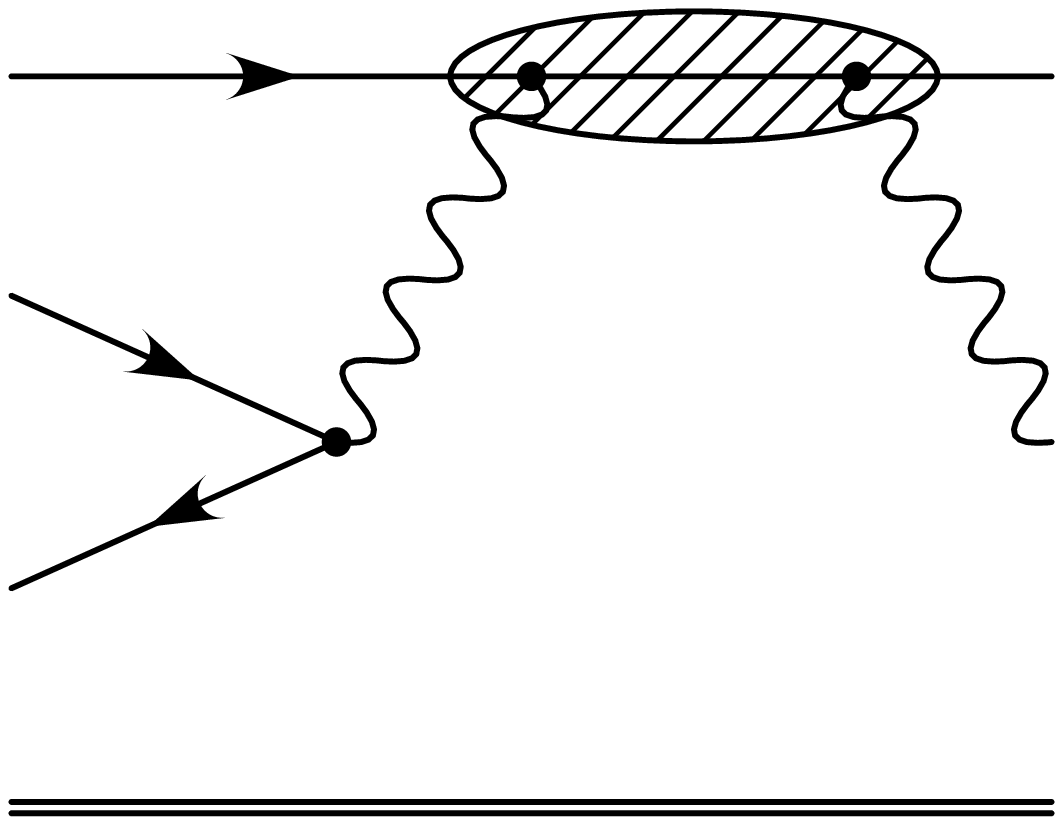}\\[2mm]

A\hspace*{50mm}B\\[8mm]

\includegraphics[scale=0.4]{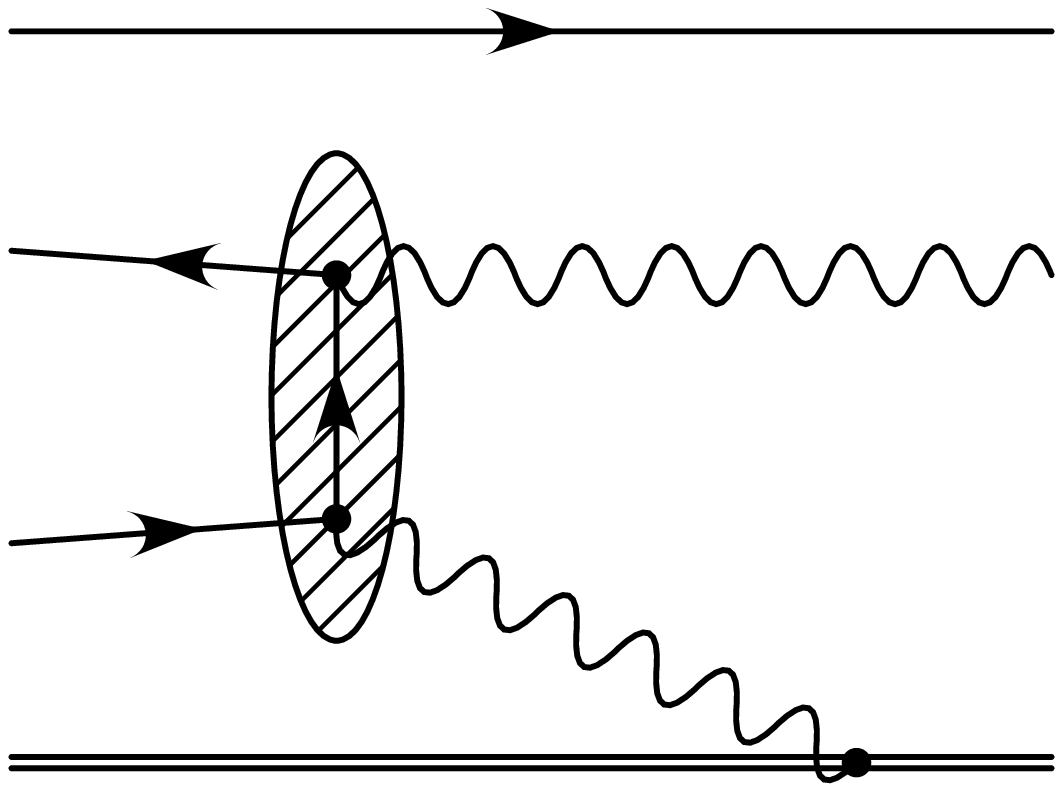}\hspace*{10mm}\includegraphics[scale=0.4]{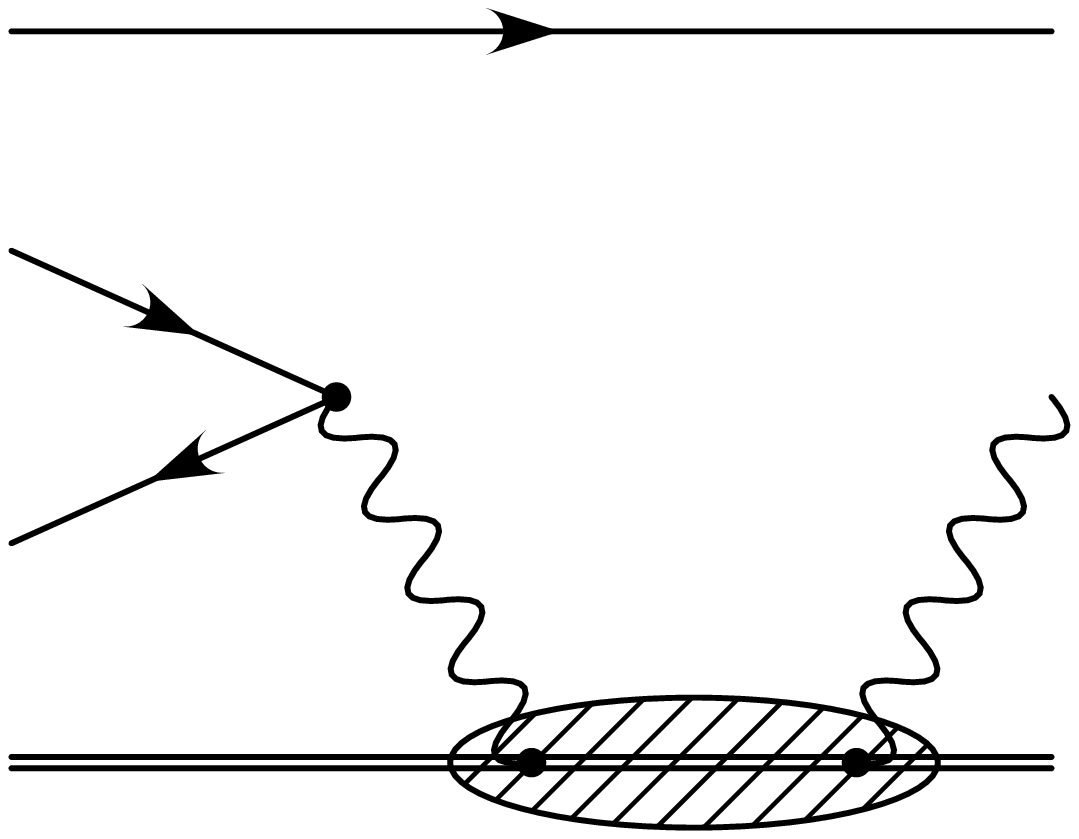}\\[2mm]

C\hspace*{50mm}D\\[8mm]

\includegraphics[scale=0.5]{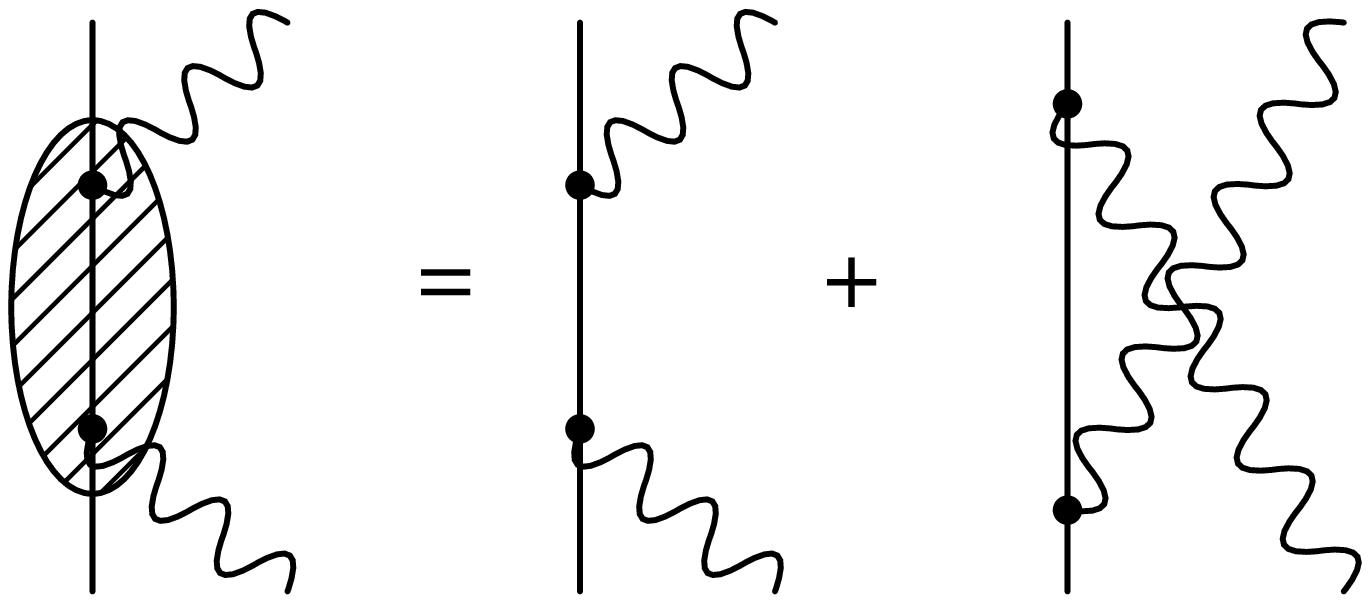}

\caption{Diagrams contributing to the decay $\text{PsH}\to p^{+}e^{-}\gamma$.
Electrons and positrons are represented by solid straight lines and
the proton by a double line. Blobs indicate two possible orderings
of photon couplings.\label{fig:DiagramsOneGamma}}
\end{figure}

Four types of diagrams can contribute to the decay $\text{PsH}\to p^{+}e^{-}\gamma$,
as shown in Fig.~\ref{fig:DiagramsOneGamma}. In all of them, an
$e^{+}e^{-}$ pair annihilates into one or two photons. One of the
produced photons is absorbed by the spectator electron or by the proton. 

We want to argue that the dominant (by far) contribution is provided
by diagrams in Fig.~\ref{fig:DiagramsOneGamma}A and B, where a photon
is absorbed by the spectator electron. Diagrams C and D are strongly
suppressed and can be neglected. Since PsH is weakly bound, its constituents' velocities
 are small and can be neglected. In that limit,
the proton can be treated as a static source of Coulomb photons. In
group C, the two-photon annihilation occurs only for a spin-singlet
$e^{+}e^{-}$ pair. The spin-singlet projector contains $\gamma^{5}$
\citep{Czarnecki:1999mw} and, for the amplitude not to vanish, Dirac
matrices $\gamma^{1,2,3}$ must be supplied by  vertices and by
electron's propagator. Interaction with a Coulomb photon,
coupled via $\gamma^{0}$, does not contribute. Similarly, in group
D, the matrix $\gamma^{0}$ has a zero matrix element between spinors
of a positron and an electron at rest. 

For this reason, it is sufficient to consider groups A and B, up to
corrections suppressed by powers of $\alpha$ which are small and
beyond the scope of this work. These two groups are the same as the
diagrams responsible for the positronium ion decay $\text{Ps}^{-}\to e^{-}\gamma$,
first evaluated in \citep{Kryuchkov94} and recently confirmed in
\citep{Aslam:2021uqu}. The only difference is in the coalescence
probability of $e^{-}e^{-}e^{+}$ which is much larger in PsH than
in the ion, thanks to the attraction of electrons to the proton.

When the ion Ps$^{-}$ is isolated, we know that it is approximately
a Ps atom accompanied by an electron far away \citep{Puchalski:2007ck}.
In the presence of a proton, this configuration becomes more compact.
If PsH resembles a hydrogen molecule, one may expect the two electrons
to be predominantly between the proton and the positron, binding the
system. It is reasonable to expect that the probability of $e^{-}e^{-}e^{+}$
coalescence to scale like the inverse volume of the system, which
we can estimate as proportional to $1/r_{e^{-}e^{-}}^{3}$ where $r_{e^{-}e^{-}}$
is the mean distance between the electrons. Using numbers in Table
\ref{tab:Values-of-physical} and those for the ion from Ref.~\citep{Frolov99},
we get the volume ratio $\left[r\left(\text{Ps}^{-}\right)/r\left(\text{PsH}\right)\right]^{3}$
equal about 13.6. This is consistent with the ratio of coalescence
probabilities: for PsH, Eq.~(\ref{eq:coal3}) gives $\left\langle \delta_{e^{+}e_{3}^{-}}\delta_{e_{3}^{-}e_{4}^{-}}\right\rangle =3.73\left(2\right)\cdot10^{-4}$,
which is about 10 times larger than $0.35875\left(2\right)\cdot10^{-4}$
in the ion $\text{Ps}^{-}$ \citep{Frolov99}. This consistency among
various estimates obtained with the variational approach is reassuring. 

Finally, we obtain the one-photon decay rate by substituting the PsH
value of $\left\langle \delta_{e^{+}e_{3}^{-}}\delta_{e_{3}^{-}e_{4}^{-}}\right\rangle $
into Kryuchkov's \citep{Kryuchkov94} result for the Ps$^{-}$,
\begin{align}
\Gamma\left(\text{PsH}\to p^{+}e^{-}\gamma\right) & =\frac{64\pi^{2}}{27}\alpha^{9}\me\left\langle \delta_{e^{+}e_{3}^{-}}\delta_{e^{+}e_{4}^{-}}\right\rangle =0.398\left(8\right)\text{s}^{-1}.\label{eq:on-photon5}
\end{align}
We have quadrupled the error arising from the numerical evaluation
of the coalescence probability to account for corrections of higher
order in $\alpha$.

\section{Zero-photon decay $\text{PsH}\to p^{+}e^{-}$\label{sec:Zero-photon-decay}}

PsH can also decay with only an electron and a proton in the final
state, $\text{PsH}\to p^{+}e^{-}$, when photons produced in the
$e^+e^-$ annihilation are absorbed by surviving components of
PsH (internal conversion). This channel is very suppressed because it
requires all four constituent to coalesce, and also it is of a higher
order in $\alpha$.  Its signature is a relativistic electron with
energy of about $3\me$. Since our result for this decay differs from
previous studies by orders of magnitude, we describe our calculation
in detail.

\begin{figure}[h]
\centering\includegraphics[scale=0.4]{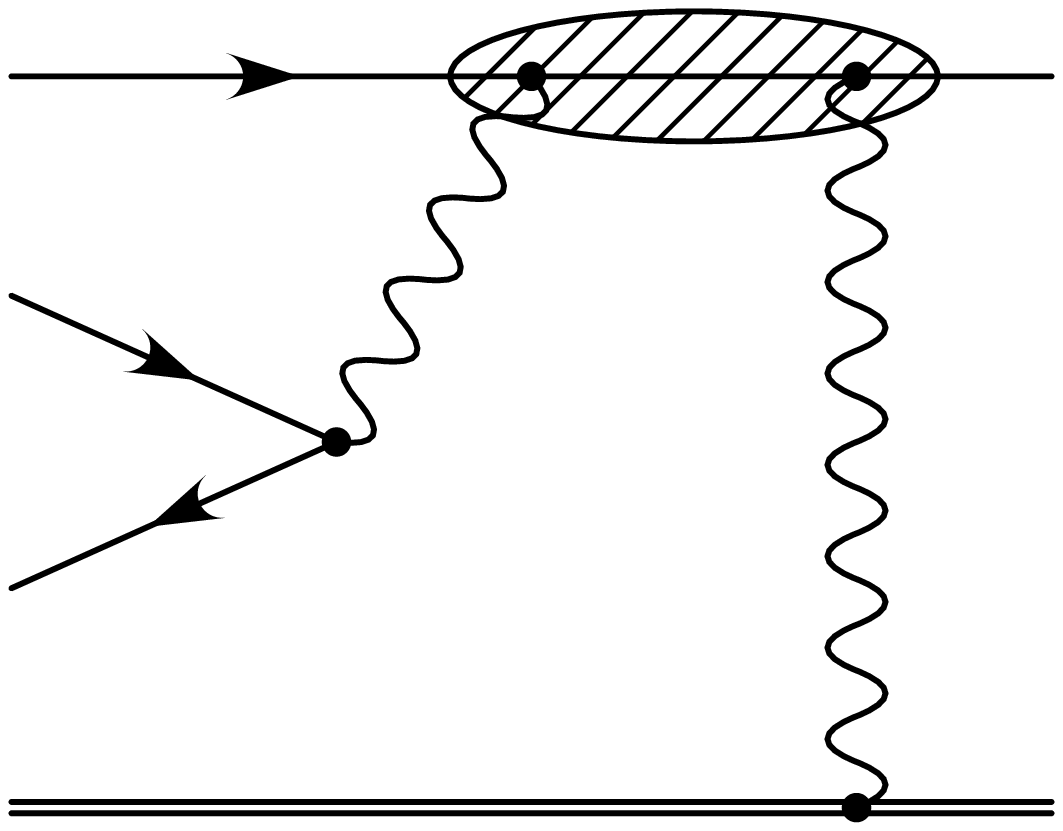}\hspace*{10mm}\includegraphics[scale=0.4]{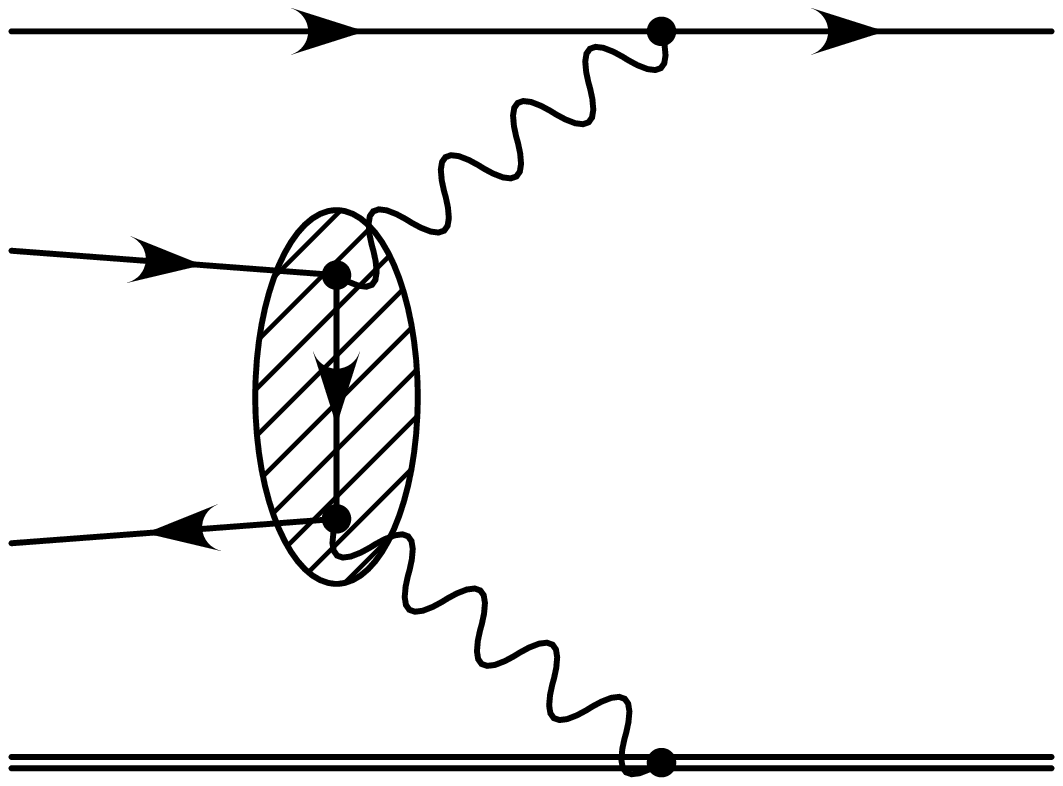}\hspace*{10mm}\includegraphics[scale=0.4]{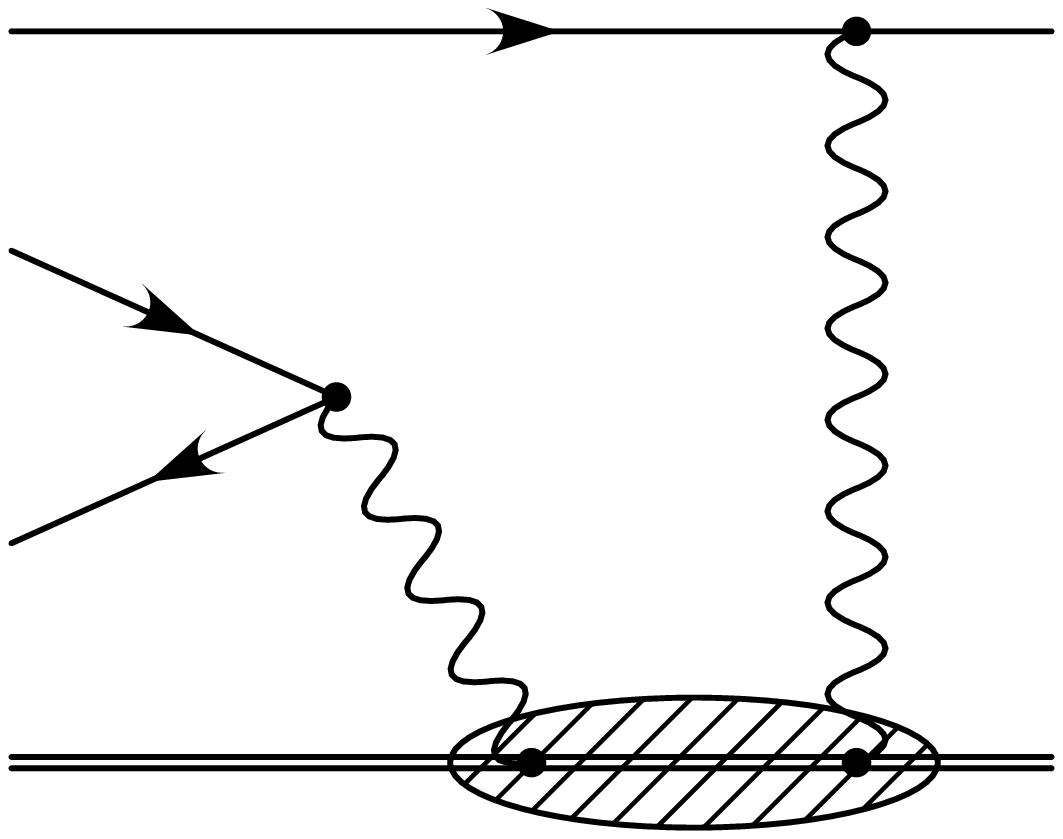}\\[2mm]

A\hspace*{50mm}B\hspace*{50mm}C

\caption{Diagrams contributing to the decay $\text{PsH}\to p^{+}e^{-}$ with
no photons in the final state. As in Fig.~\ref{fig:DiagramsOneGamma},
blobs denote two  orderings of photon couplings.\label{fig:DiagramsNoGamma}}
\end{figure}
Diagrams contributing to the decay $\text{PsH}\to p^{+}e^{-}$ are
shown in Fig.~\ref{fig:DiagramsNoGamma}. They are divided into three
groups A, B, C, differing by the topology of the photon exchange.
Working in the leading order in the velocities of the constituent
particles, one can neglect groups B and C, by the same reasoning as
at the beginning of Section \ref{sec:One-photon-decay}.

\begin{figure}[h]
\centering\includegraphics[scale=0.4]{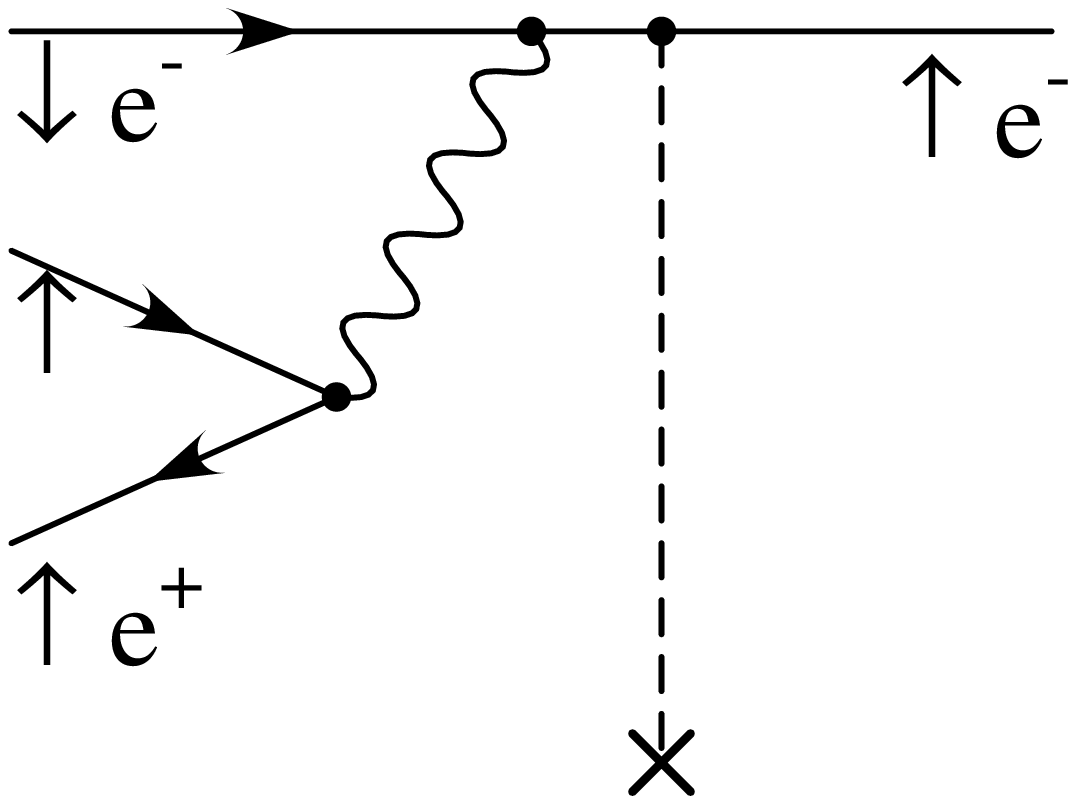}\hspace*{10mm}\includegraphics[scale=0.4]{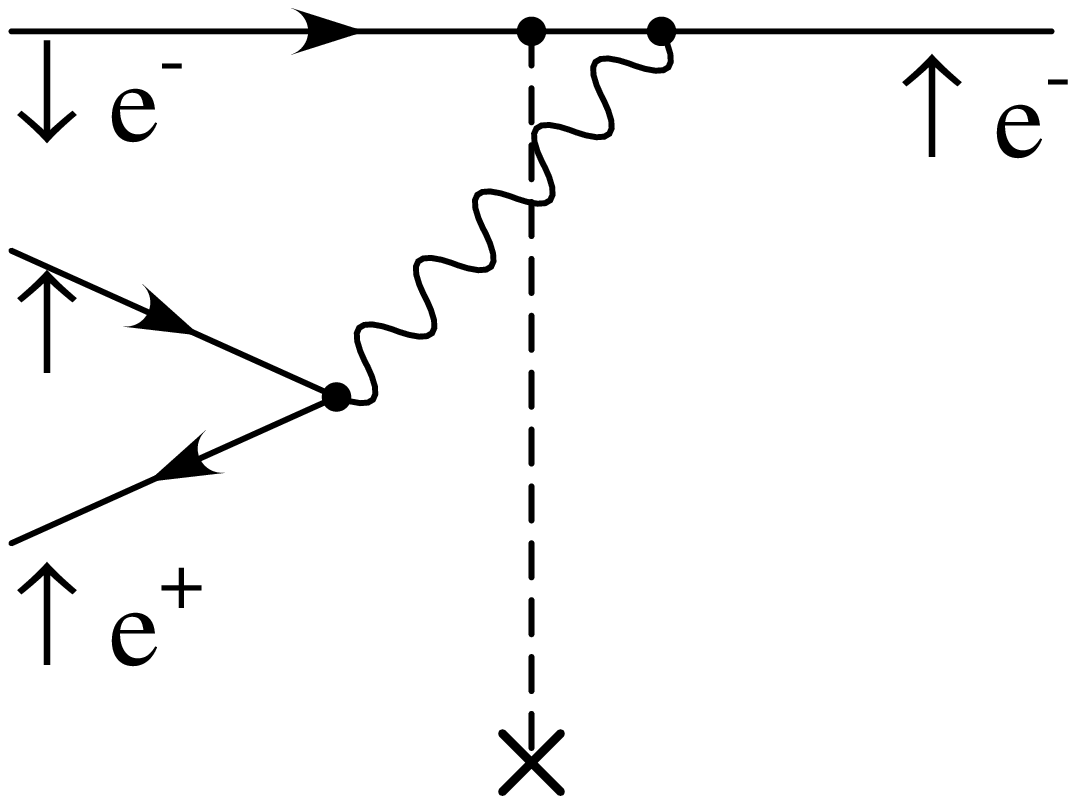}\\[2mm]

A1\hspace*{50mm}A2\\[2mm]

\includegraphics[scale=0.4]{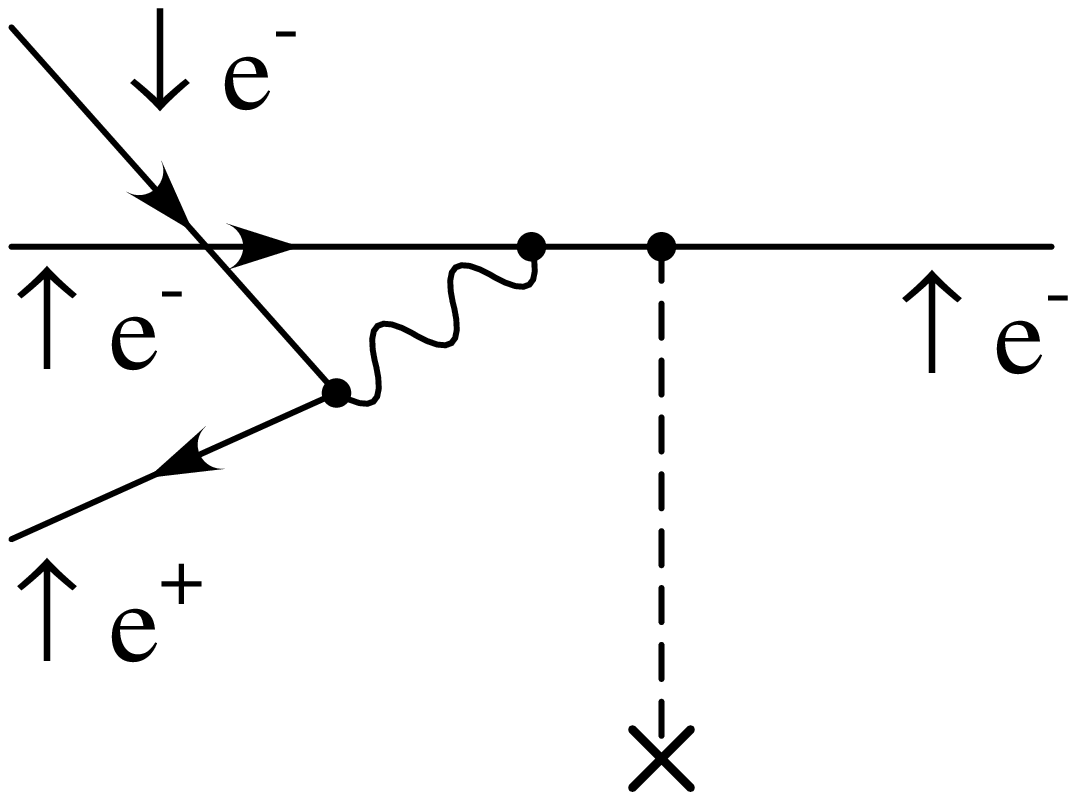}\hspace*{10mm}\includegraphics[scale=0.4]{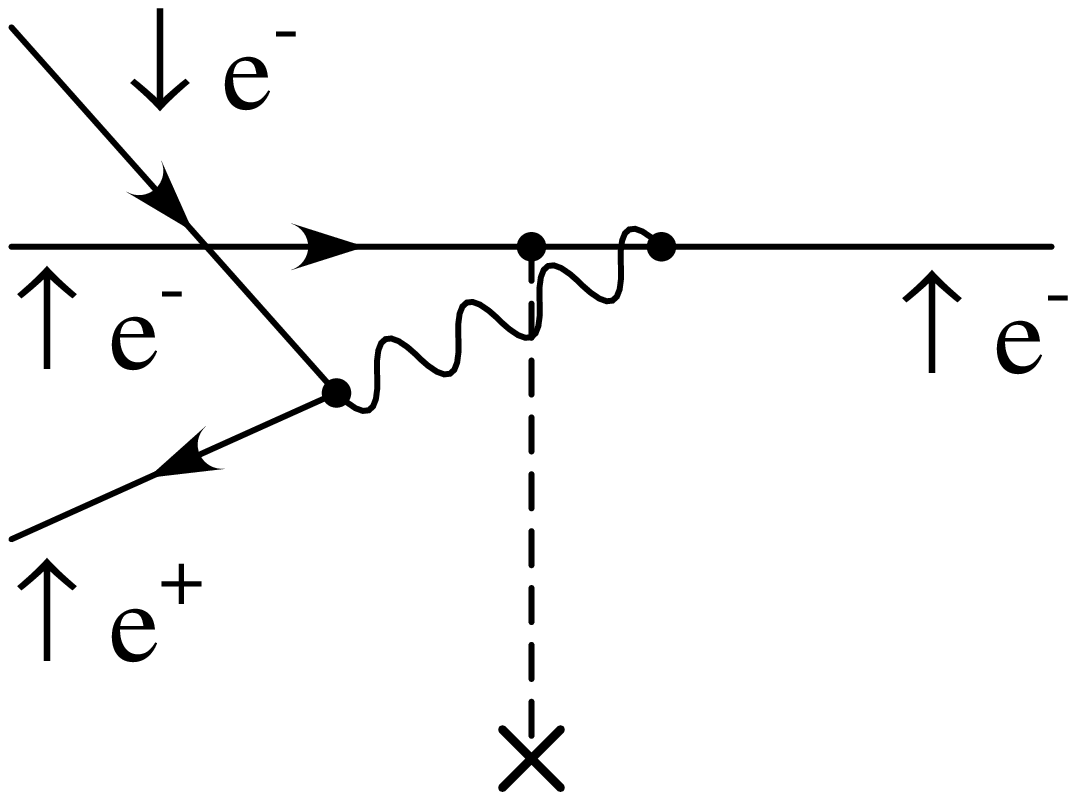}\\[2mm]

A3\hspace*{50mm}A4\\[2mm]

\caption{Diagrams of group A for the decay $\text{PsH}\to p^{+}e^{-}$ in the
limit of an infinitely massive proton. Dashed line denotes interaction
with the Coulomb field of the proton.\label{fig:GroupA}}
\end{figure}

We therefore evaluate only diagrams in group A, shown in
Fig.~\ref{fig:GroupA}.  We frame the calculation as a decay of $\text{Ps}^-$ in an external Coulomb field. Choosing the
$z$ axis along the polarization of the positron, we compute the
amplitude of the electron emission along that axis. The electron
emitted in that direction must be right-handed since it carries the
spin of the initial state. The amplitude of emission at a non-zero
polar angle $\theta$ will be multiplied by
$\cos\left(\theta/2\right)$, resulting in a factor
$\left\langle \cos^{2}\left(\theta/2\right)\right\rangle =1/2$ in the
decay rate.  That factor is canceled when the rate of decay into a
left-handed electron is included. (If daughter electrons' polarization
is not observed, their angular distribution is isotropic because of
$\cos^{2}\left(\theta/2\right)+\sin^{2}\left(\theta/2\right)=1$.)

The daughter electron carries the rest energy of the initial state,
$E_{f}=3\me$. For it to be on the mass shell, the Coulomb photon
exchanged with the nucleus (dashed line in Fig.~\ref{fig:GroupA})
must carry momentum $p_{f}=2\sqrt{2}\me$ in the $z$ direction. Its
propagator supplies factor of  $e/\left(8\me^{2}\right)$ to the amplitude.
The remaining factors for the amplitudes pictured in Fig.~\ref{fig:GroupA}
are (amplitudes 3 and 4 contain a minus sign relative to 1, 2, due
to permutation of fermion operators)
\begin{align}
\calM_{1} & =\calM_{2}=-\frac{e^{3}}{8\sqrt{2}m^{3}},\quad\calM_{3}=-\frac{e^{3}}{16\sqrt{2}m^{3}},\quad\calM_{4}=\frac{3e^{3}}{16\sqrt{2}m^{3}},\\
\calM & =\sqrt{2}\left(\calM_{1}+\calM_{2}+\calM_{3}+\calM_{4}\right)=-\frac{e^{3}}{8m^{3}},
\end{align}
where the factor of  $\sqrt{2}$ arises from the electron spin singlet
wave function, $\left(\uparrow\downarrow-\downarrow\uparrow\right)/\sqrt{2}$:
there are two equal contributions divided by $\sqrt{2}$. These results
are obtained assuming free particles annihilating at rest, using the
daughter electron's spinor $u_{f}^{\dagger}=\left(\begin{array}{cccc}
1 & 0 & 1/\sqrt{2} & 0\end{array}\right)$. In order to account for the binding, the amplitude is multiplied
\citep{Aslam:2021uqu} by the PsH wave function at zero separation
among the positron and electrons. The square of the amplitude is summed
over the final states. The rate is a product of four factors: final-state
normalization, amplitude squared, phase space, and the coalescence
probability that includes $1/2!$ accounting for identical electrons,
\begin{align}
\Gamma\left(\text{PsH}\to p^{+}e^{-}\right) & =\frac{1}{u_{f}^{\dagger}u_{f}}\cdot\left(\frac{e^{4}}{64m^{5}}\right)^{2}\cdot\frac{4\pi p_{f}E_{f}}{\left(2\pi\right)^{2}}\cdot\frac{\left(\alpha m\right)^{9}\left\langle \delta_{p+--}\right\rangle }{2!}\\
 & =\frac{\sqrt{2}}{8}\pi^{3}\alpha^{13}\left\langle \delta_{p+--}\right\rangle m,\label{eq:PsHNoPhotons}
\end{align}
where $\left\langle \delta_{p+--}\right\rangle $ denotes $\left\langle \delta_{p^{+}e^{+}}\delta_{p^{+}e_{3}^{-}}\delta_{e^{+}e_{4}^{-}}\right\rangle =1.85\left(1\right)\cdot10^{-4}$
given in Eq.~(\ref{eq:coal4}). Using this value we get the rate
\begin{equation}
\Gamma\left(\text{PsH}\to p^{+}e^{-}\right)=1.31\left(7\right)\cdot10^{-10}\text{ s}^{-1}.\label{eq:noPhoton}
\end{equation}

\section{Conclusions\label{sec:Conclusions}}

We have determined rates of two rare decays of the ground state of
positronium hydride and confirmed a number of basic properties for
this system using the variational principle with a Gaussian basis. 

In the case of one photon annihilation, $\text{PsH}\to p^{+}e^{-}\gamma$,
where one of the photons produced in the $e^{+}e^{-}$ annihilation
can be absorbed either by the electron or by the proton, we have demonstrated
that the proton contribution is negligible. When the electron absorbs
the photon, the decay resembles that of the already extensively studied $\text{Ps}^{-}$ ion. We find (see Eq.~(\ref{eq:on-photon5}))
\begin{equation}
\Gamma\left(\text{PsH}\to p^{+}e^{-}\gamma\right)=0.398\left(8\right)\text{s}^{-1}.
\end{equation}
The assigned error includes the spread of values of the coalescence
probability, higher order $\alpha$ corrections, and much smaller
proton recoil effects. This result should be compared with previous
estimates. Ref.~\citep{Frolov:1997hy}  assumed (incorrectly) that the contribution
of the photon absorption by the proton ``does not differ significantly
from'' that by the electron and thus obtained a rate about twice
larger than we did, $0.8077\,\text{s}^{-1}$ (Table V in Ref.~\citep{Frolov:1997hy}).
Similarly, Ref.~\citep{Bubin06} repeated the claim that absorptions
by the electron and by the proton contribute approximately equally
and obtained $0.787501\,\text{s}^{-1}$ using a slightly different
coalescence probability. We stress, once again, that the photon absorption
by the proton is suppressed by the velocity of constituents of PsH,
equivalent to a suppression by $\alpha$.

The other decay channel we considered was the radiationless decay
$\text{PsH}\to p^{+}e^{-}$ for whose rate we found in Eq.~(\ref{eq:noPhoton})
$1.31\left(7\right)\cdot10^{-10}\text{ s}^{-1}$.
The previous estimate \citep{Frolov:1997hy}, $9.16\times10^{-8}\text{ s}^{-1}$,
is larger by a factor of almost 700. That estimate was obtained by
using the dipositronium 
$\text{Ps}_{2}$ result (Eq.~(32) in Ref.~\citep{Frolov:1997hy}).
There are two problems with this reasoning. First, the $\text{Ps}_{2}$
formula used in Ref.~\citep{Frolov:1997hy} was incorrect even for
$\text{Ps}_{2}$: it overestimated the zero-photon decay rate of $\text{Ps}_{2}$
by a factor of about 5.44 \citep{Aslam:2021uqu}. What about the remaining
factor of $700/5.44\simeq130$? The $\text{Ps}_{2}$ decay is quite
different from that of PsH. The numerical coefficient in $\text{Ps}_{2}$
is $27\sqrt{3}/2\simeq23$ \citep{Aslam:2021uqu} instead of that
in PsH being $\sqrt{2}/8\simeq0.18$ (see our Eq.~(\ref{eq:PsHNoPhotons})).
Their ratio is $23/0.18\simeq130$, explaining the remaining discrepancy. 

This large ratio has several sources: different symmetry factors,
the proton not contributing in PsH, and, crucially, different particle
virtualities. In the PsH decay, the emitted electron carries a large
momentum with a magnitude of $\sqrt{8}m$. The propagator of the Coulomb
photon supplying this momentum introduces a large suppression factor.
Just to illustrate how this leads to large numbers, consider the diagram
similar to Fig.~\ref{fig:DiagramsNoGamma}B in the decay $\text{Ps}_{2}\to e^{+}e^{-}$:
the denominators in the propagators of the photons and of the virtual
electron are, in units of $1/m^{2}$, $-1/2,-1/2,1/4$, producing
$1/16$. Now consider denominators in Fig.~\ref{fig:DiagramsNoGamma}A
for $\text{PsH}\to p^{+}e^{-}$: $1/4,1/8,-1/8$, giving $-1/256$.
Rates involve squares of these products, favoring the $\text{Ps}_{2}$
rate by the relative factor of 256. This illustrates how the ratio
of 130 of the $\text{Ps}_{2}$ and PsH rates is quite natural. 
\begin{acknowledgments}
We thank Professor Yasuyuki Nagashima for bringing positronium hydride
decays to our attention. 
We thank Bo Leng and Kyle McKee for collaboration at an early stage of this project.
Our variational code is a modified version of a program we received
from Mariusz Puchalski during earlier studies \cite{Puchalski:2007ck,Puchalski:2008jj}.
This research was supported by Natural Sciences and Engineering Research
Council of Canada (NSERC).
\end{acknowledgments}


\begin{thebibliography}{10}
\providecommand{\url}[1]{\texttt{#1}}
\providecommand{\urlprefix}{URL }
\providecommand{\eprint}[2][]{\url{#2}}

\bibitem{Bubin11}
S.~Bubin and K.~Varga, \emph{Ground-state energy and relativistic corrections
  for positronium hydride}, Phys. Rev. A \textbf{84}, 012509 (2011).

\bibitem{Wheeler:1946xth}
J.~A. Wheeler, \emph{Polyelectrons}, Ann. N. Y. Acad. Sci. \textbf{48},
  219--238 (1946).

\bibitem{Hylleraas:1947zza}
E.~A. Hylleraas and A.~Ore, \emph{Binding Energy of the Positronium Molecule},
  Phys. Rev. \textbf{71}, 493--496 (1947).

\bibitem{Ore:1951zz}
A.~Ore, \emph{{The Existence of Wheeler-Compounds}}, Phys. Rev. \textbf{83},
  665--665 (1951).

\bibitem{Neamtan1962}
S.~M. Neamtan, G.~Darewych, and G.~Oczkowski, \emph{{Annihilation of Positrons
  from the ${\mathrm{H}}^{\ensuremath{-}}{e}^{+}$ Ground State}}, Phys. Rev.
  \textbf{126}, 193--196 (1962).

\bibitem{goldanskii:1967aa}
V.~I. Goldanskii, \emph{{The Quenching of Positronium and the Inhibition of its
  Formation}}, in A.~T. Stewart and L.~O. Roellig, editors, \emph{{Positron
  Annihilation: Proc. of the Conference Held at Wayne State University, July
  27-29, 1965}}, pages 183--258 (1967).

\bibitem{Lebeda:1969}
C.~F. Lebeda and D.~M. Schrader, \emph{Towards an Accurate Wave Function for
  Positronium Hydride}, Phys. Rev. \textbf{178}, 24--34 (1969).

\bibitem{Navin1974}
P.~B. Navin, D.~M. Schrader, and C.~F. Lebeda, \emph{An improved wave function
  for positronium hydride: Preliminary report}, Appl. Phys. \textbf{3},
  159--160 (1974).

\bibitem{Ho:1978}
Y.~K. Ho, \emph{A resonant state and the ground state of positronium hydride},
  Phys. Rev. A \textbf{17}, 1675--1678 (1978).

\bibitem{Yoshida:1996}
T.~Yoshida and G.~Miyako, \emph{{Diffusion quantum Monte Carlo calculations of
  positronium hydride and positron lithium}}, Phys. Rev. A \textbf{54},
  4571--4572 (1996).

\bibitem{Frolov:1997aa}
A.~M. Frolov and V.~H. Smith, \emph{Ground state of positronium hydride}, Phys.
  Rev. A \textbf{56}, 2417--2420 (1997).

\bibitem{Nagashima:1998}
Y.~Nagashima, T.~Hyodo, K.~Fujiwara, and A.~Ichimura, \emph{Momentum-transfer
  cross section for slow positronium-He scattering}, Journal of Physics B:
  Atomic, Molecular and Optical Physics \textbf{31}, 329--339 (1998).

\bibitem{Pareja90}
R.~Pareja, R.~M. de~la Cruz, M.~A. Pedrosa, R.~Gonz\'alez, and Y.~Chen,
  \emph{{Positronium hydride in hydrogen-laden thermochemically reduced MgO
  single crystals}}, Phys. Rev. B \textbf{41}, 6220--6226 (1990).

\bibitem{Schrader92}
D.~M. Schrader, F.~M. Jacobsen, N.-P. Frandsen, and U.~Mikkelsen,
  \emph{Formation of positronium hydride}, Phys. Rev. Lett. \textbf{69}, 57--60
  (1992).

\bibitem{Yan99}
Z.-C. Yan and Y.~K. Ho, \emph{Relativistic effects in positronium hydride},
  Phys. Rev. A \textbf{60}, 5098--5100 (1999).

\bibitem{Houston73}
S.~K. Houston and R.~J. Drachman, \emph{Comment on the Ground State of
  Positronium Hydride}, Phys. Rev. A \textbf{7}, 819--820 (1973).

\bibitem{Page1974}
B.~A.~P. Page and P.~A. Fraser, \emph{The ground state of positronium hydride},
  Journal of Physics B: Atomic and Molecular Physics \textbf{7}, L389--L392
  (1974).

\bibitem{NavinPRA74}
P.~B. Navin, D.~M. Schrader, and C.~F. Lebeda, \emph{Improved wave function for
  positronium hydride}, Phys. Rev. A \textbf{9}, 2248--2251 (1974).

\bibitem{Ferrante:1968zz}
G.~Ferrante, \emph{{Annihilation of Positrons from Positronium Negative Ion
  $e^-e^+e^-$}}, Phys. Rev. \textbf{170}, 76--80 (1968).

\bibitem{Frolov:1997hy}
A.~M. Frolov and V.~H. Smith, \emph{{Positronium hydrides and the
  ${\mathrm{Ps}}_{2}$ molecule: Bound-state properties, positron annihilation
  rates, and hyperfine structure}}, Phys. Rev. A \textbf{55}, 2662--2673
  (1997).

\bibitem{FROLOV2005430}
A.~M. Frolov, \emph{{Positron annihilation in the positronium negative ion
  Ps$^-$}}, Physics Letters A \textbf{342}, 430 -- 438 (2005).

\bibitem{Frolov:2009qi}
A.~M. Frolov, \emph{{Annihilation of electron-positron pairs in the positronium
  ion Ps$^-$ and bipositronium Ps$_2$}}, Phys.~Rev.~A \textbf{80}, 014502
  (2009).

\bibitem{Puchalski:2008jj}
M.~Puchalski and A.~Czarnecki, \emph{{Dipole Excitation of Dipositronium}},
  Phys. Rev. Lett. \textbf{101}, 183001 (2008), \eprint{0810.0013}.

\bibitem{Ho1986}
Y.~K. Ho, \emph{Positron annihilation in positronium hydrides}, Phys. Rev. A
  \textbf{34}, 609--611 (1986).

\bibitem{Czarnecki:1999mw}
A.~Czarnecki, K.~Melnikov, and A.~Yelkhovsky, \emph{{Positronium S state
  spectrum: Analytic results at ${\mathcal O}(m \alpha^6)$}}, Phys. Rev.
  \textbf{A59}, 4316 (1999), \eprint{hep-ph/9901394}.

\bibitem{Kryuchkov94}
S.~I. Kryuchkov, \emph{{On the one-photon annihilation of the Ps$^-$ ion}}, J.
  Phys. B \textbf{27}, L61 (1994).

\bibitem{Aslam:2021uqu}
M.~J. Aslam, W.~Chen, A.~Czarnecki, S.~R. Mir, and M.~Mubasher, \emph{{Rare
  decays of the positronium ion and molecule, $\text{Ps}^{-}\to e^{-}\gamma$
  and $\text{Ps}_{2}\to e^{+}e^{-}\gamma,~\gamma\gamma,~e^{+}e^{-}$}}, Phys.
  Rev. A \textbf{104}, 052803 (2021), \eprint{2108.06785}.

\bibitem{Puchalski:2007ck}
M.~Puchalski, A.~Czarnecki, and S.~G. Karshenboim, \emph{{Positronium-ion
  decay}}, Phys. Rev. Lett. \textbf{99}, 203401 (2007), \eprint{0711.0008}.

\bibitem{Frolov99}
A.~M. Frolov, \emph{Bound-state properties of the positronium negative ion
  $Ps^{-}$}, Phys. Rev. A \textbf{60}, 2834--2839 (1999).

\bibitem{Bubin06}
S.~Bubin and L.~Adamowicz, \emph{Nonrelativistic variational calculations of
  the positronium molecule and the positronium hydride}, Phys. Rev. A
  \textbf{74}, 052502 (2006).

\end{thebibliography}

\end{document}